\documentclass{emulateapj}
\usepackage{graphicx}

\slugcomment {}

\shorttitle{Discovery of a High Latitude Supernova Remnant}

\shortauthors{Fesen et al.}

\begin{document}

\title{Discovery of an Apparent High Latitude Galactic Supernova Remnant}

\author{Robert A.\ Fesen, Jack M.\ M.\ Neustadt, Christine S.\ Black, \& Ari H.\ D.\ Koeppel}
\affil{6127 Wilder Lab, Department of Physics \& Astronomy, Dartmouth
  College, Hanover, NH 03755 USA}

\begin{abstract}

Deep H$\alpha$ images of a faint emission complex $4.0\degr \times 5.5\degr$ in
angular extent and located far off the Galactic plane at $l = 70.0\degr$,
$b=-21.5\degr$ reveal numerous thin filaments suggestive of a supernova
remnant's shock emission. Low dispersion optical spectra covering the
wavelength range 4500 - 7500 \AA \ show only Balmer line emissions for one
filament while three others show a Balmer dominated spectrum along with weak
[\ion{N}{1}] 5198, 5200 \AA, [\ion{O}{1}] 6300, 6364 \AA, [\ion{N}{2}] 6583
\AA, [\ion{S}{2}] 6716, 6731 \AA \ and in one case [\ion{O}{3}] 5007 \AA \ line
emission.  Many of the brighter H$\alpha$ filaments are visible in near UV {\sl
GALEX} images presumably due to \ion{C}{3}] 1909 \AA \ line emission. {\sl
ROSAT} All Sky Survey images of this region show a faint crescent shaped X-ray
emission nebula coincident with the portion of the H$\alpha$ nebulosity closest
to the Galactic plane.  The presence of long, thin Balmer dominated emission
filaments with associated UV emission and coincident X-ray emission suggests
this nebula is a high latitude Galactic supernova remnant despite a lack of
known associated nonthermal radio emission. Relative line intensities of the
optical lines in some filaments differ from commonly observed
[\ion{S}{2}]/H$\alpha$ $\geq 0.4$ radiative shocked filaments and typical
Balmer filaments in supernova remnants.  We discuss possible causes for the
unusual optical SNR spectra.

\end{abstract}

\keywords{ISM: individual objects: G70.0-21.5, ISM: supernova remnant - 
 shock waves - optical - X-rays: ISM}

\section{Introduction}

Currently, there are 294 confirmed and nearly two dozen possible or suspected
Galactic supernova remnants (SNRs) \citep{Green14}.  The majority of these were
first identified in the radio due to nonthermal radio emissions associated with
shocked gas \citep{Milne70,Downes71}.  Roughly 40\% of the known Galactic
SNRs exhibit associated X-ray emission with a smaller percentage ($\sim$30\%)
showing some coincident optical emission.

New Galactic SNRs are still occasionally identified in the radio
\citep{deGasp14,Gao14,Gerbrandt14} and X-rays \citep{Renaud10,Reynolds13}.
Although discoveries of Galactic remnants in the optical are relatively rare,
several have recently been made (e.g., \citealt{SP08,FM10,Sabin13}) through
deep H$\alpha$ surveys such as the H$\alpha$ emission Virginia Tech Spectral
Line Survey (VTSS) of the Galactic Plane \citep{Dennison98,Fink03} and the
Isaac Newton Telescope Photometric H$\alpha$ Survey \citep{Drew05,GS08}.
 
Here we report the discovery of an apparent Galactic SNR located well off the
Galactic plane.  The remnant consists of faint emission covering a region some
$4.0\degr \times 5.5\degr$ in angular size and spherical in shape along its
eastern boundary.  Deep H$\alpha$ imaging and optical spectroscopic
observations of portions of this emission complex are described in $\S$2 with
the results presented and discussed in $\S$3.

\section{Observations}

\begin{figure*}[t]
\centering
\includegraphics[scale=0.91]{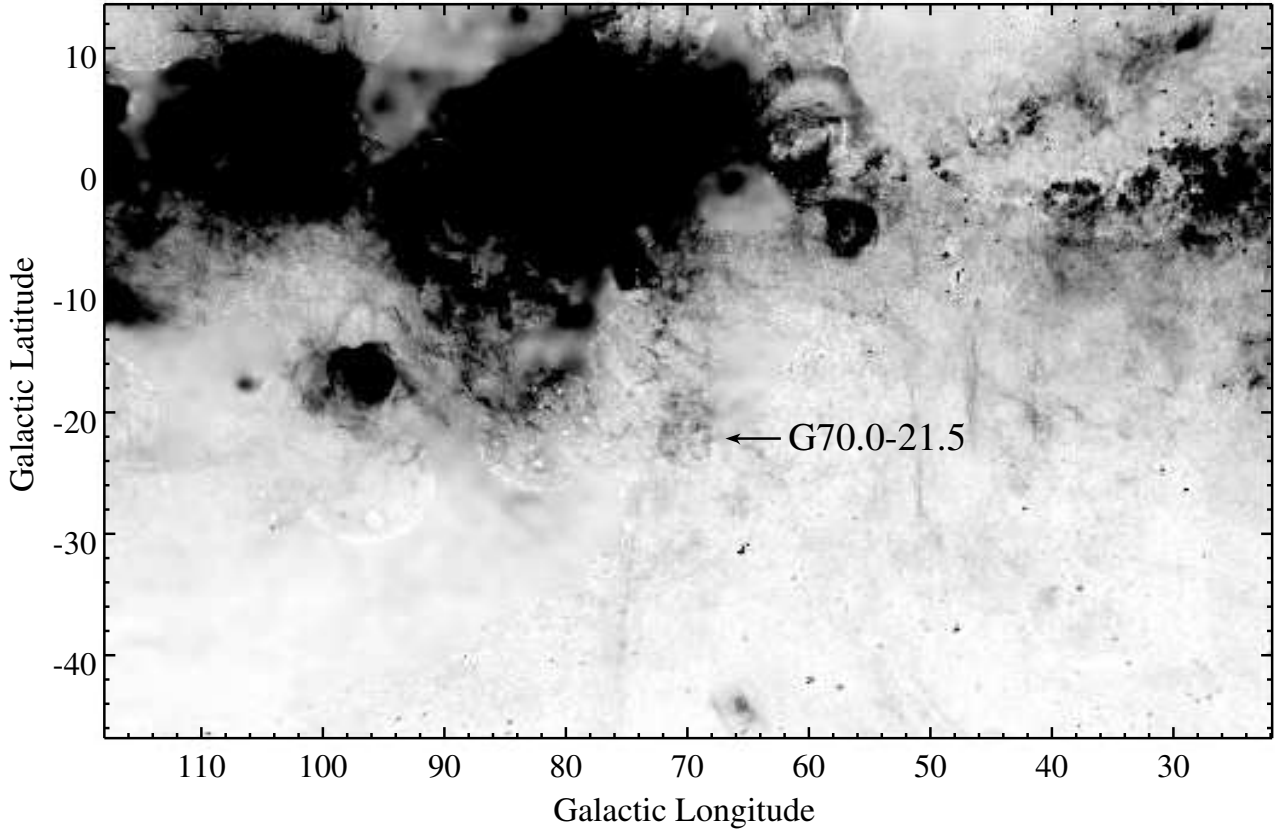}
\caption{H$\alpha$ image of the G70.0-21.5 nebulosity as seen on the Virginia
Tech Spectral Line Survey (VTSS) of the Galactic Plane.
}
\label{VTSS_wide_view}
\end{figure*}

Examination of VTSS H$\alpha$ images revealed a large, faint, and relatively
isolated optical emission nebulosity located well off the galactic plane.
Figure \ref{VTSS_wide_view} shows this nebula in the VTSS
H$\alpha$ survey.  The nebulosity is roughly $4.0\degr \times 5.5\degr$ in
angular extent and centered approximately at $l = 70.0\degr$, $b=-21.5\degr$
corresponding to RA = $21^{\rm h}$ $24^{\rm m}$, Dec = $+19\degr$ $23'$
(J2000).

This emission is not part of a cataloged galactic nebula or H II region
\citep{Sharpless59,Mar74,Neckel85}. It is also too faint to be seen in the
H$\alpha$ survey by \citet{Sivan74} and lies too far off the plane to be
covered in the optical emission line survey of the Milky Way by
\citet{Parker79}.  However, some of its brighter regions are faintly visible on
both first and second digital Palomar Observatory Sky Survey (POSS) red images.

The nebula, denoted G70.0-21.5 and hereafter also referred to as G70, exhibits
an irregular morphology but is roughly circular along its eastern boundary.
While its extent in the west is uncertain due to
missing VTSS data to the southwest and possible confusion with fainter
neighboring emission features to the west, G70's angular size is at least
$4.0\degr \times 5.5\degr$ based on its appearance on the VTSS images and our
follow-up H$\alpha$ images described below.

\begin{figure*}[bt]
\centering
\includegraphics[scale=0.75]{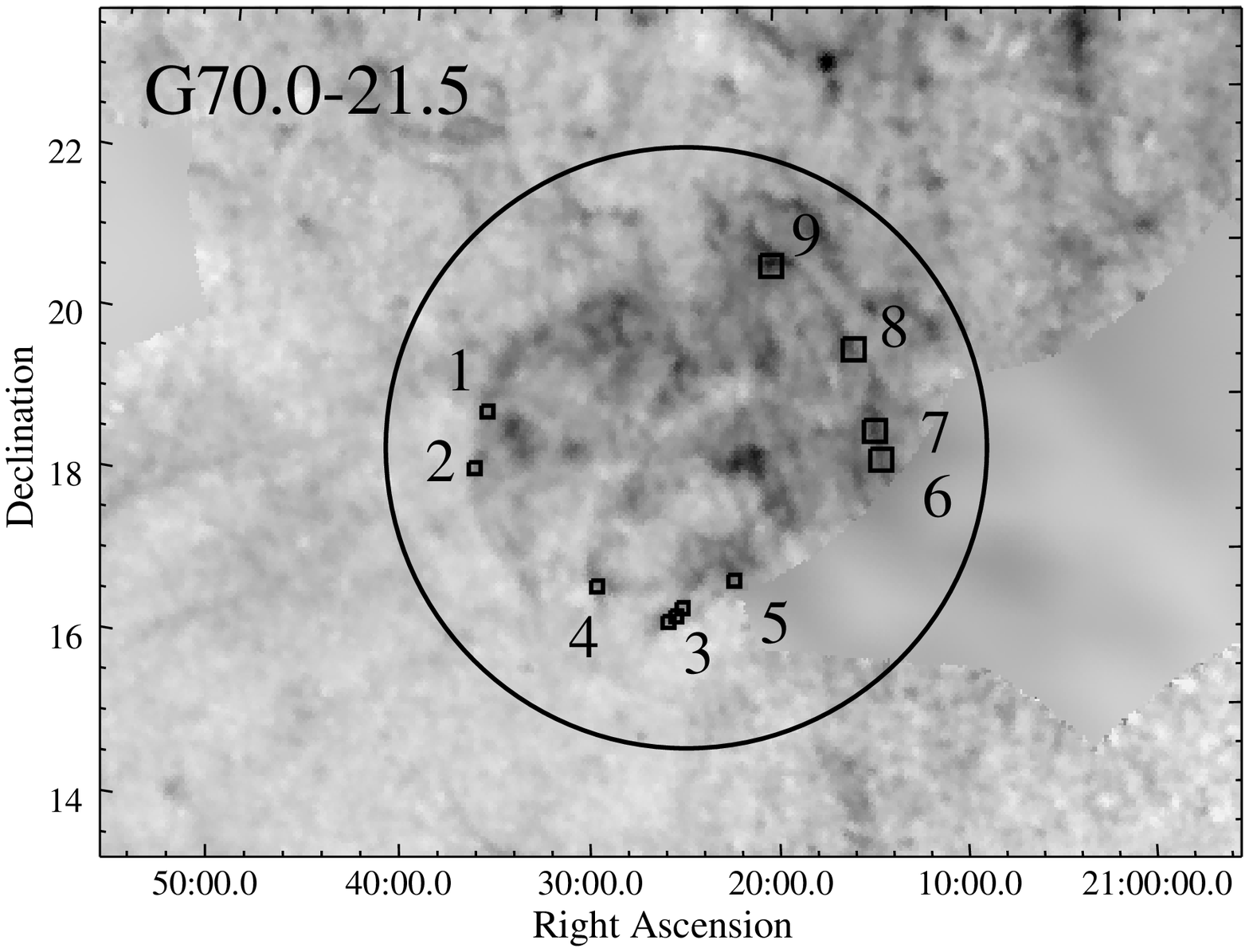}
\caption{H$\alpha$ image of G70.0-21.5 from the  Virginia
Tech Spectral Line Survey (VTSS) of the Galactic Plane.
North is up, East to the left with J2000 coordinates shown. 
The circle encloses the suspected extent of G70.0-21.5 with
the numbered boxes marking the approximate
regions ($1 - 9$) where we obtained higher resolution H$\alpha$ images.
Low dispersion spectra were taken in Regions 3 and 5; see Fig.\ 6 }
\label{VTSS}
\end{figure*}

Nine regions of this nebulosity were imaged in June and October 2014 at the MDM
Observatory at Kitt Peak (see Fig.\ \ref{VTSS}).  The images were obtained
using a backside illuminated $2048 \times 2048$ SITe CCD detector and 30 \AA \
and 90 \AA \ FWHM H$\alpha$ filters.  Regions 1 through 5 were imaged using the
2.4m Hiltner telescope which provided a $9'$ field of view with an image scale
of $0\farcs275$ pixel$^{-1}$.  Regions 6 through 9 were imaged on the 1.3m
McGraw-Hill telescope producing a field of view of approximately $17'$ with an
image scale of $0\farcs508$ pixel$^{-1}$.  Table 1 lists the coordinate centers
and exposure times for all nine regions.

Low-dispersion optical spectra of filamentary features in this nebula
were obtained in June 2014 using the MDM 2.4~m telescope with the Mark III
Spectrograph and a 1024 $\times$ 1024 Tektronix CCD detector.  A $1\farcs 2
\times 4.5'$ slit and a 300 lines mm$^{-1}$ $5400$ \AA \ blaze grism were used
to obtain sets of two or three $1000 - 1200$ s exposures spanning the spectral
region $4500 - 7500$ \AA\ with a spectral resolution of 8 \AA.
 
Standard pipeline data reduction of these images and spectra was performed
using IRAF\footnote{IRAF is distributed by the National Optical Astronomy
Observatories, which is operated by the Association of Universities for
Research in Astronomy, Inc.\ (AURA) under cooperative agreement with the
National Science Foundation.}. This included debiasing, flat-fielding using
twilight sky flats, and dark frame corrections.  Spectral data were reduced
using standard IRAF software routines and calibrated with Hg, Ne, and Xe lamps
and \citet{Oke74} \citet{Massey90} standard stars. Flat fielding of the images
did not always result in flat background intensity levels for some of the 2.4 m
images, most apparent along the edges of the images for Regions 1 and 2.

\section{Results and Discussion}

\subsection{H$\alpha$ Images}

Although appearing diffuse in the low resolution (1\farcm6 pixel$^{-1}$) VTSS
images with only a hint of filamentary structure from its relatively sharp
eastern edge, our H$\alpha$ images of nine regions of this nebulosity 
(Fig.\ \ref{VTSS}) revealed dozens of thin emission filaments (Figs.\ 3 -- 5).
The H$\alpha$ filaments in Regions 1, 2, 3 and 5
along the nebula's eastern and southern edges indicate a definite boundary to
the emission complex in these areas.  The presence of long, thin filaments in
the western Regions 6 through 9, like those detected in the east, suggests the
nebula extends at least out to these locations.

\begin{figure*}
\epsscale{1.0}
\centering
\plottwo{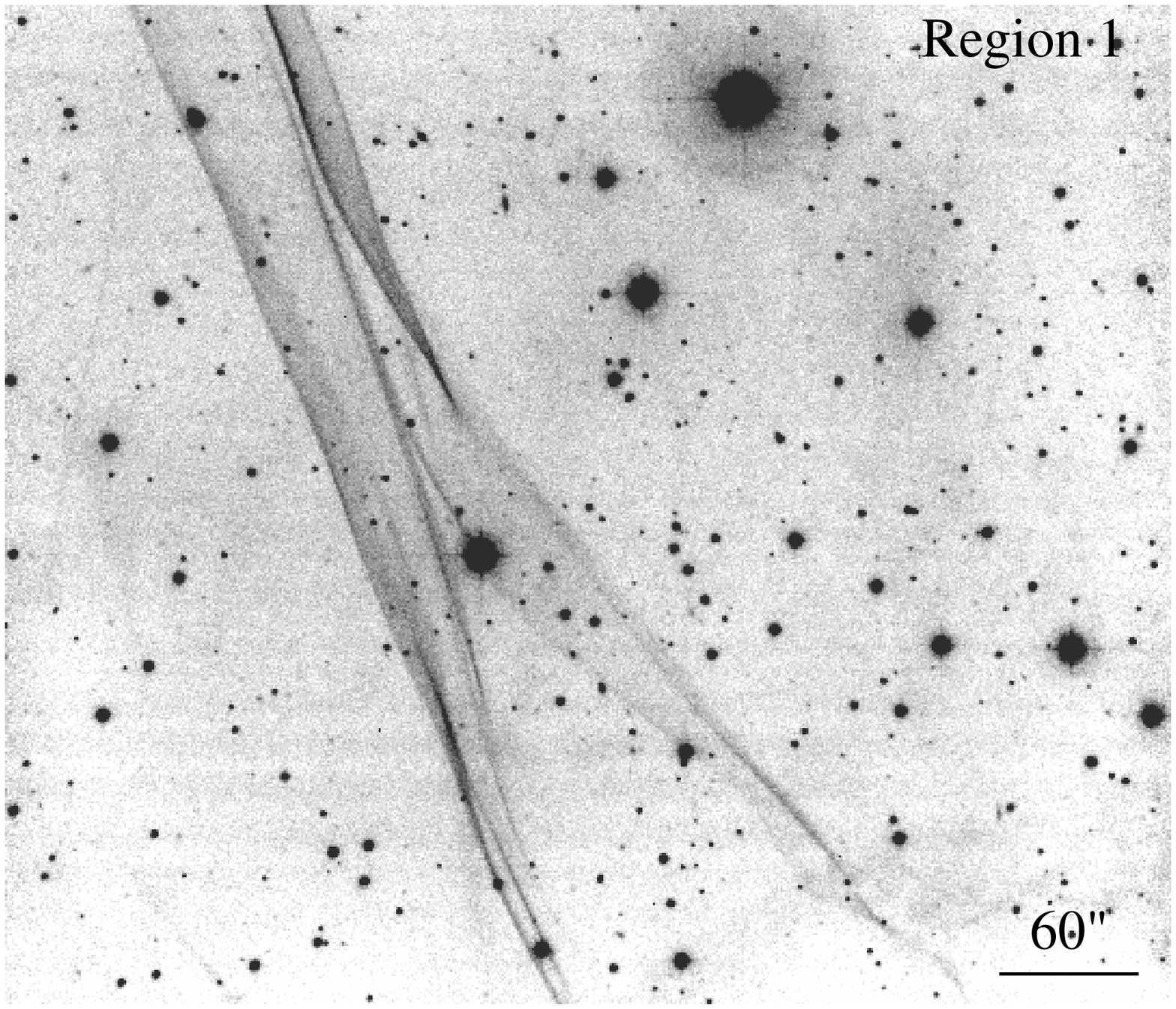}{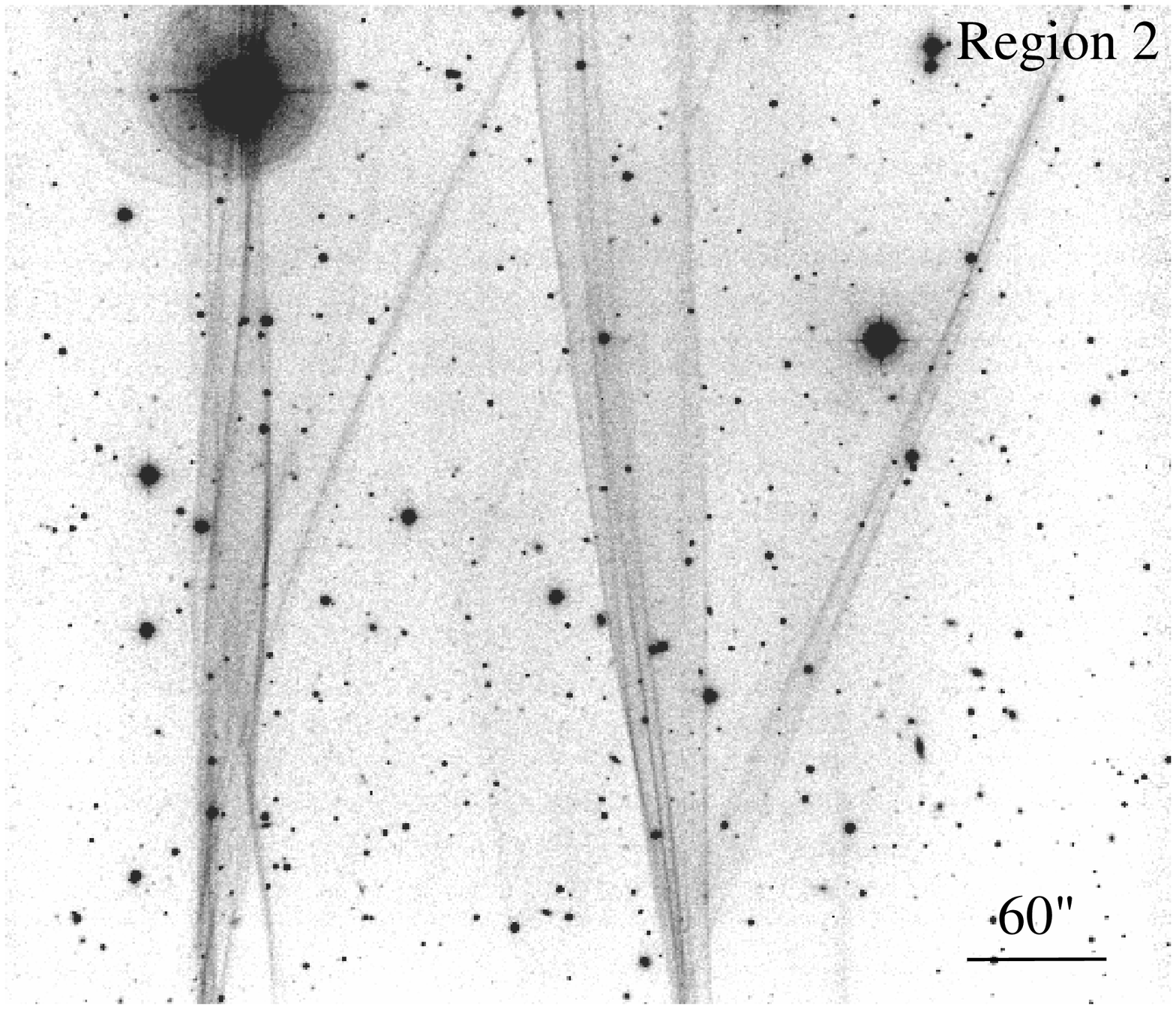}  \\
\plottwo{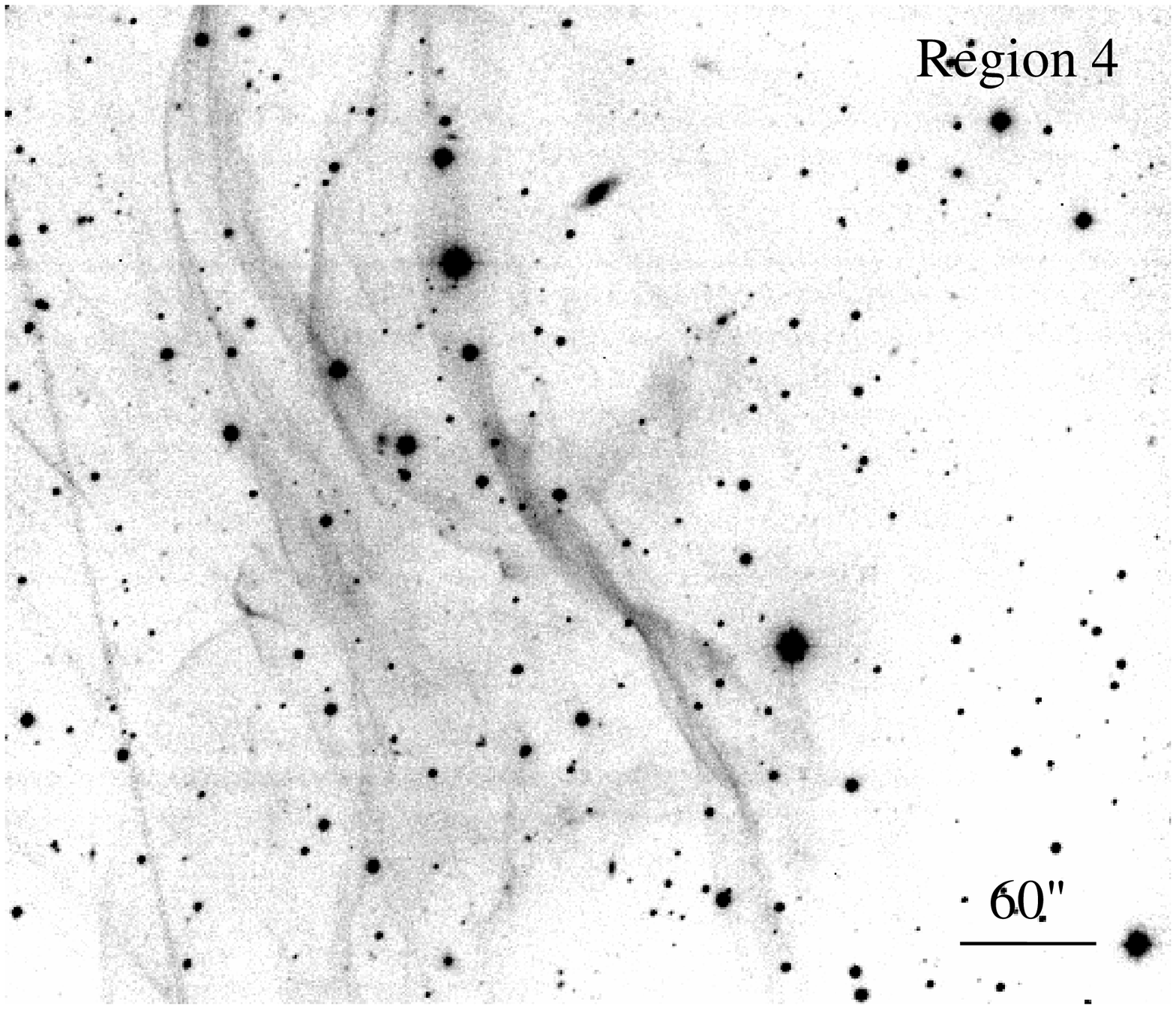}{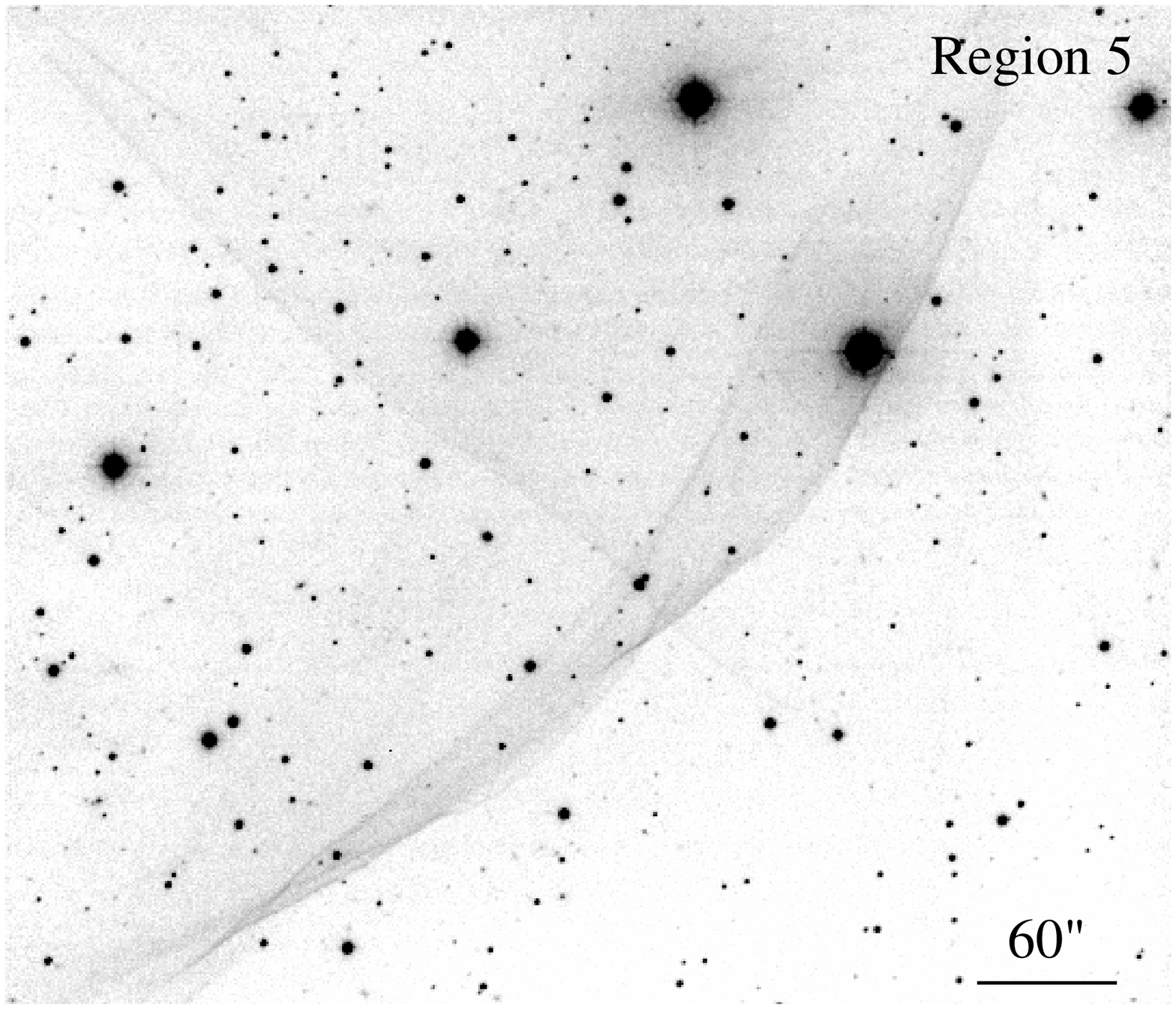}  \\
\caption{H$\alpha$ images of Regions 1, 2, 4, and 5, in G70.0-21.5.  North is up, East to the left.
}
\label{Reg1n5}
\end{figure*}

\begin{figure*}
\epsscale{1.0}
\centering
\plottwo{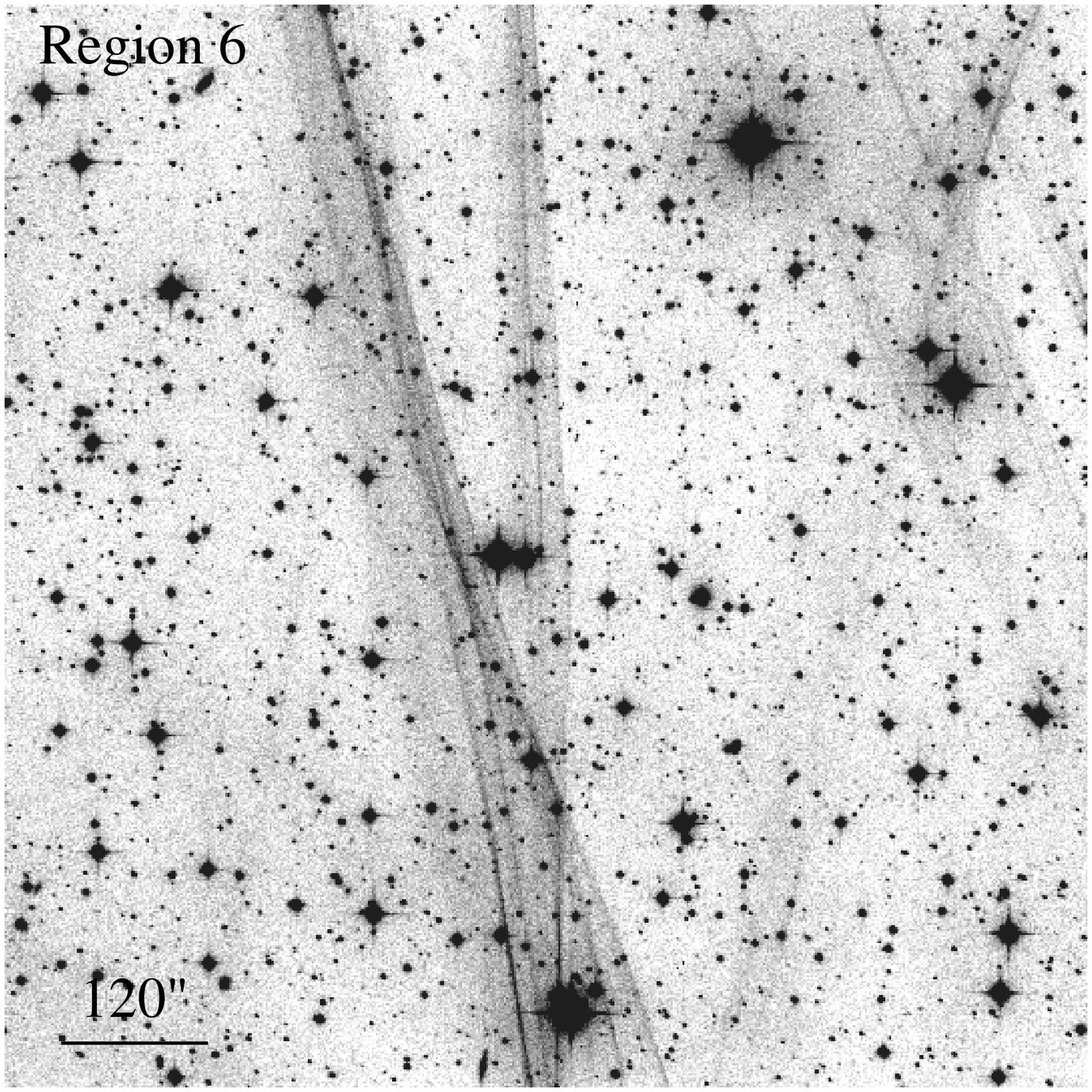}{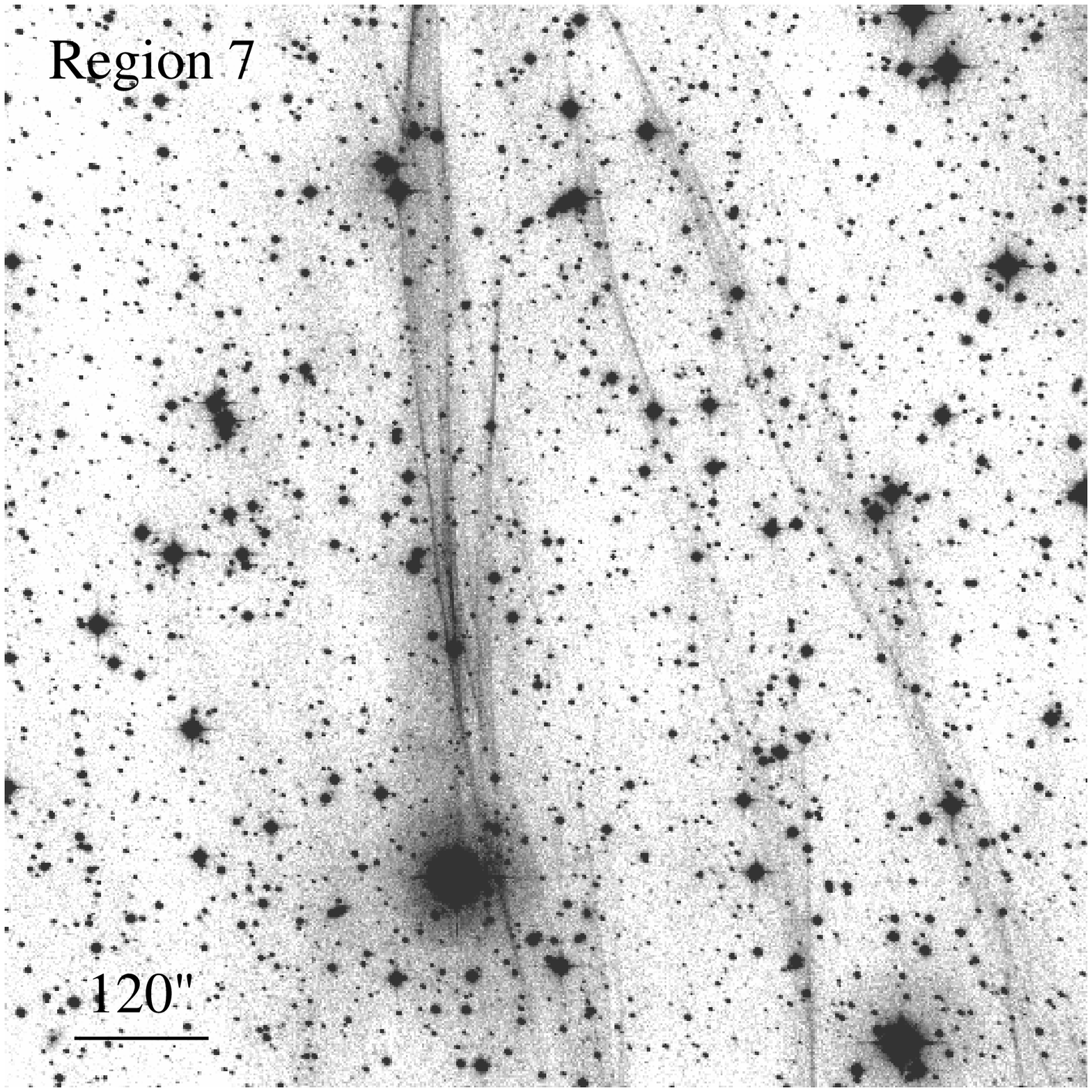} \\
\plottwo{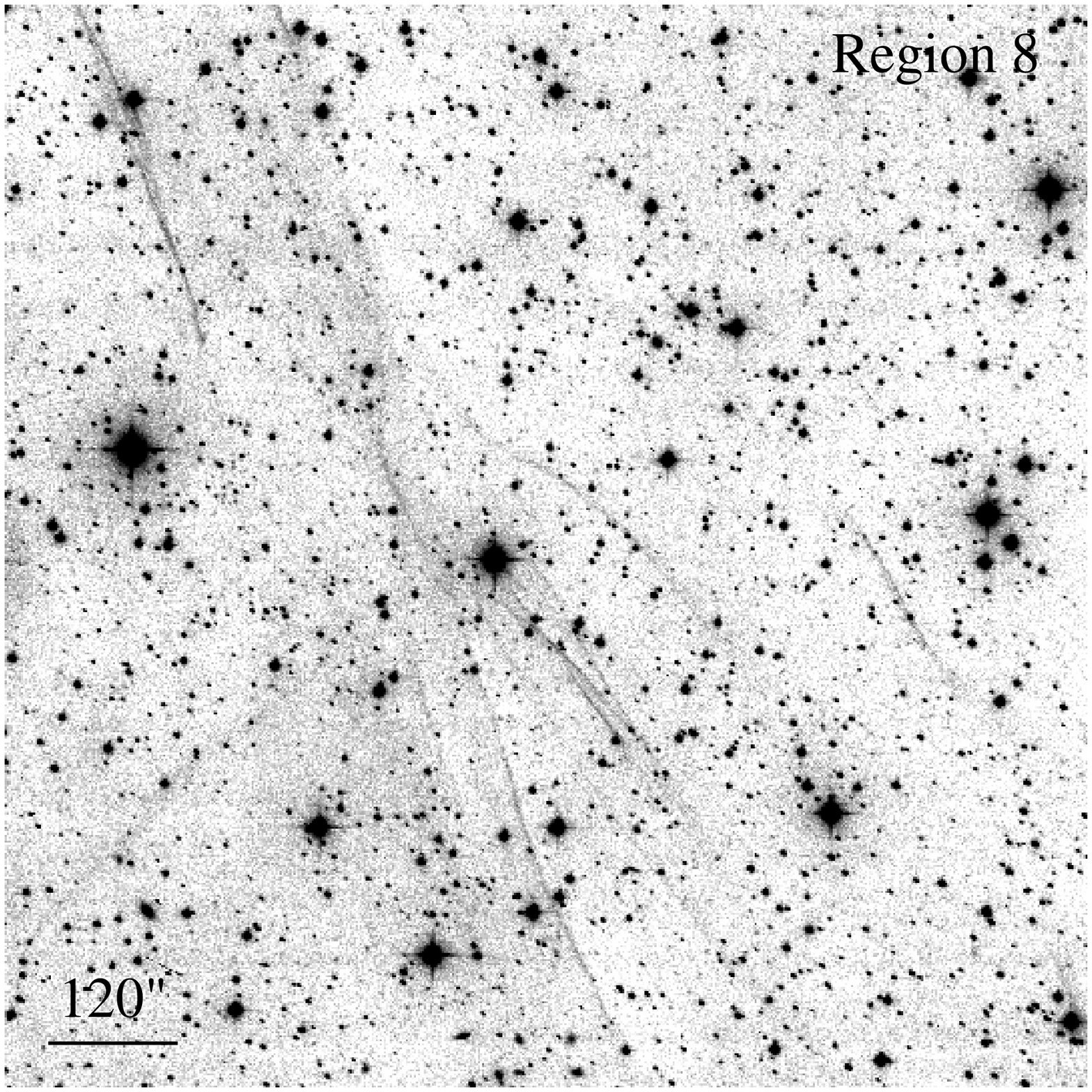}{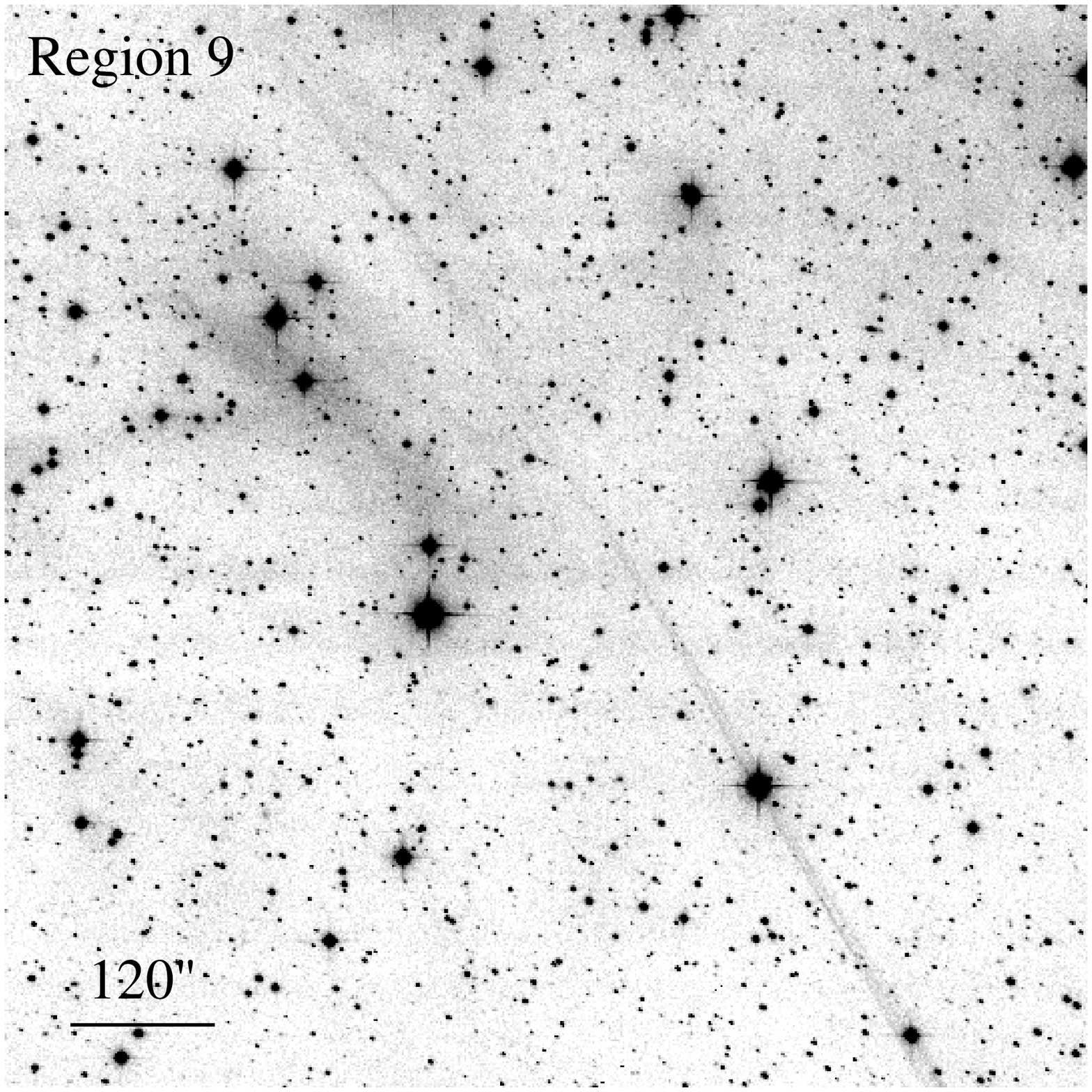}
\caption{H$\alpha$ images of Regions 6, 7, 8 and 9 in G70.0-21.5.  North is up, East to the left.}
\label{Reg6n9}
\end{figure*}

\begin{figure*}
\centering
\includegraphics[scale=0.9]{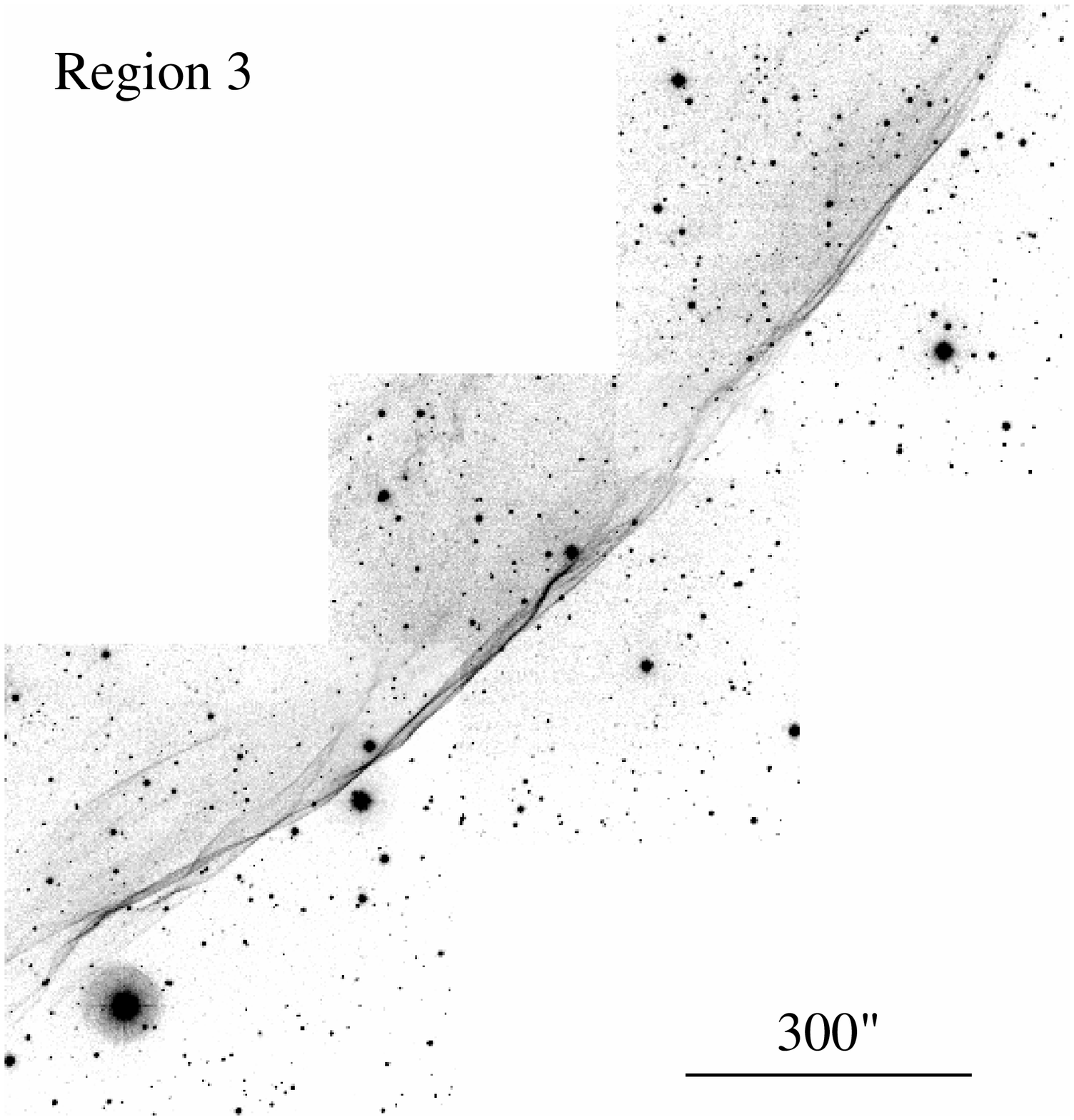}
\caption{Mosaic of H$\alpha$ images covering Region 3 in G70.0-21.5.  North is up, East to the left.
}
\label{Reg3}
\end{figure*}

Faint and unresolved filaments appear throughout much of the G70 nebula,
forming a complex of overlapping filamentary structures in places, often
curved, and extending many arcminutes in length.  An example is in Region 3
(Fig.\ \ref{Reg3}) where a long continuous emission filament lies along the
southern border of the nebula. 

The close spacing and alignment of filaments, as seen in Regions 1, 2, 6, and
7, is suggestive of a thin expanding emission shell being viewed nearly edge on
where small velocity differences across its surface create the appearance of
close or overlapping filaments.  Consistent with this picture,
faint diffuse trailing emission can be seen associated with many filaments. 

While much of G70's nebulosity is filamentary, it also contains patches
of diffuse emission, especially along its western boundary.  Our 
image of Region 9 shows considerable diffuse emission which extends to the
northwestern edge of the image frame, and an exploratory image taken farther
north showed only a broad band of diffuse emission. 

\subsection{Optical Spectra}

\begin{figure*}[t]
\epsscale{1.07}
\centering
\plotone{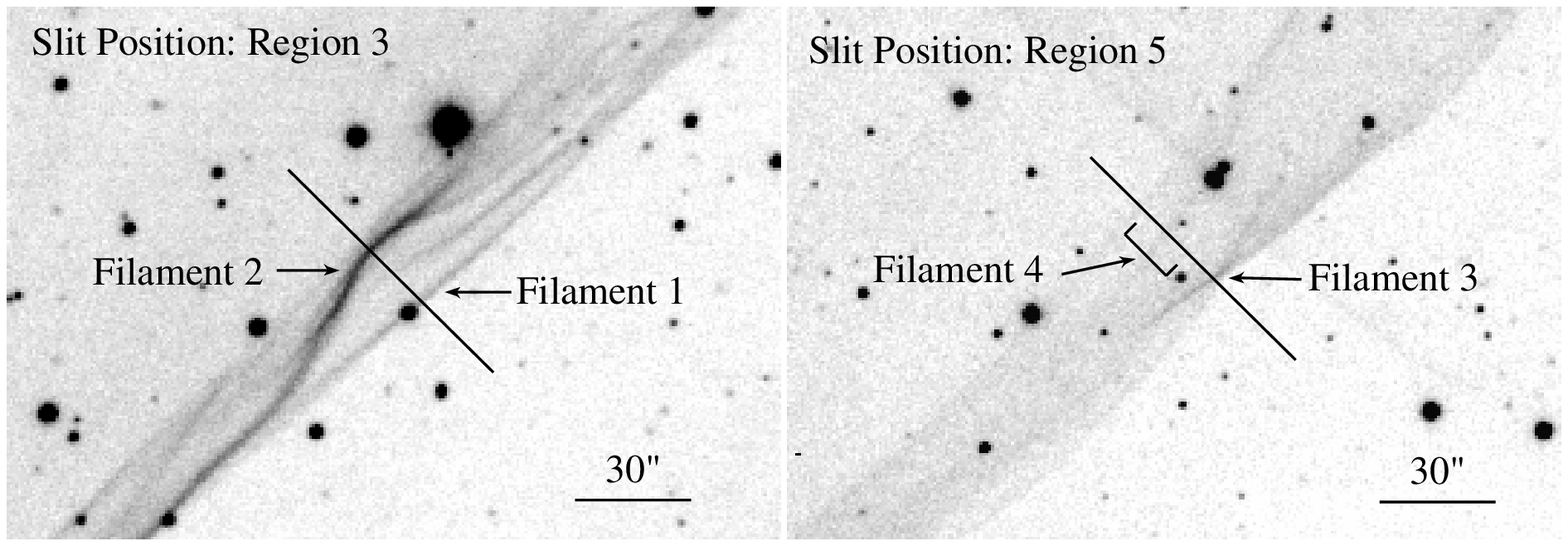}
\caption{Slit locations on H$\alpha$ images 
of Regions 3 and 5 in G70.0-21.5 where spectra where taken.
North is up, East to the left.}
\label{slits}
\end{figure*}

\begin{figure*}[t]
\centering
\includegraphics[scale=0.70,angle=270]{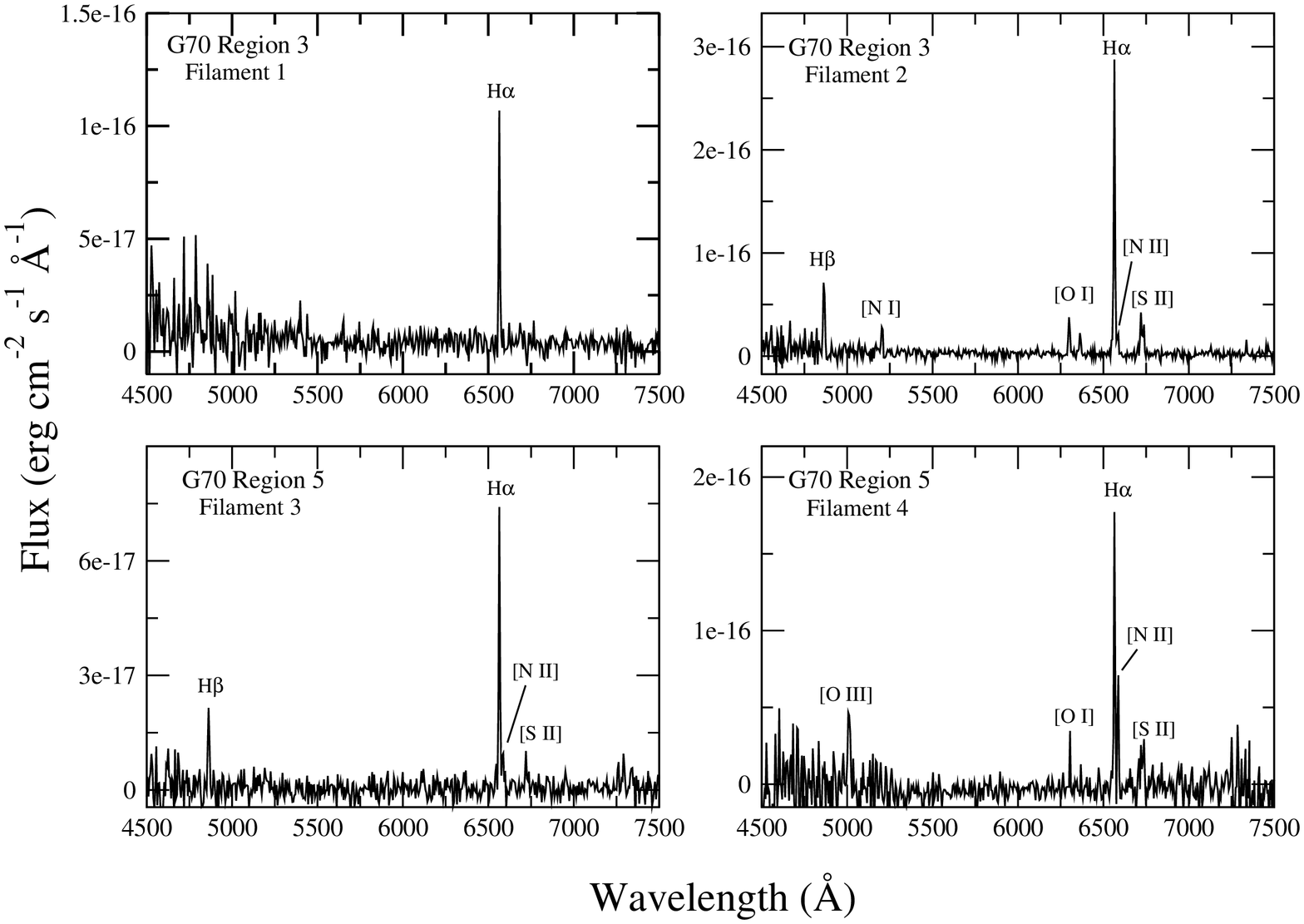}
\caption{Spectra of emission filaments in Regions 3 and 5. }
\label{spectra}
\end{figure*}

Low dispersion spectra were taken of filaments in Regions 3 and 5 with the slit
positions shown in Figure \ref{slits}.  The resulting spectra are presented in
Figure \ref{spectra}.

For Region 3, spectra were extracted for a faint, more westerly filament
(Filament 1) and for a brighter, curved filament (Filament 2) farther to the
east.  For Region 5, we extracted spectra for a small sharp filament (Filament
3) and combined a 20$''$ long region of emission behind this filament into one
spectrum (Filament 4).

Filament 1 in Region 3 was found to exhibit a Balmer dominated emission
spectrum. Only H$\alpha$ emission was clearly detected (H$\beta$ was present but
barely above the noise level) with no hint of [\ion{N}{2}] 6583 \AA \ or
[\ion{S}{2}] 6716, 6731 \AA \ line emissions down to a level below 10\% of the
strength of H$\alpha$.  This Balmer dominated type spectrum could be seen
extending some 15$''$  behind (eastward) of this filament.  

In contrast, a very different spectrum was seen for the much brighter filament
marked Filament 2 in Region 3 (see Fig.\ \ref{slits}). While this filament
also showed a Balmer dominated spectrum, it exhibited faint emission lines of
[\ion{N}{1}] 5198, 5200 \AA, [\ion{O}{1}] 6300, 6364 \AA, [\ion{N}{2}] 6583
\AA, and [\ion{S}{2}] 6716, 6731 \AA. The change in spectra between Filaments 1
and 2 was abrupt and distinct.
 
Spectra taken in Region 5 showed some aspects similar to the spectra seen for Region
3.  As shown in Figure \ref{spectra}, Filament 3's spectrum showed faint
emission from [\ion{N}{2}] 6583 \AA, and [\ion{S}{2}] 6716, 6731 \AA \ like  
that seen for Filament 2 in Region 3 but now with no [\ion{N}{1}] or [\ion{O}{1}]
emission.  Notably however, the easternmost edge of this filament, covering roughly
the first 2$''$ along the slit, showed only H$\alpha$ and H$\beta$ emission lines. 

Emission from extended nebulosity farther to the east (Filament 4) 
showed stronger [\ion{N}{2}] emission along with the additional presence of
[\ion{O}{3}] 5007 \AA \ line emission. The appearance of [\ion{O}{3}] emission
started abruptly roughly 10$''$ east of Filament 3 and extended 
some 20$''$ along the slit to the east. 

The measured H$\alpha$/H$\beta$ ratios for Filament 2 in Region 3 and Filament
3 in Region 5 are 3.14 and 3.4, respectively.  Assuming an intrinsic
H$\alpha$/H$\beta$ of 2.87 under the assumption of large optical depths in the
Lyman series (i.e., Case B; see Table $4.2$ in \citealt{Oster06}) and an
electron temperature of 10$^{4}$ K, we find a E(B--V) values of $0.085 \pm
0.02$ mag and  $0.15 \pm 0.05$ mag for Filaments 2 and 3, respectively.  Due to
the faintness of Filament 3 of Region 5 compared to Filament 2 in Region 3, we
have given more weight to its E(B--V) value and estimate an E(B-V) value for
G70 of $0.10 \pm 0.3$.

This value is close to the E(B--V) value of
$0.08$ seen for the neighboring Cygnus Loop SNR \citep{Parker67,Fesen82}. This
small amount of reddening to the G70 filaments is not surprising given their
location so far off the galactic plane. 

The electron density sensitive [\ion{S}{2}] 6716/6731 emission line ratio of
$1.3 \pm 0.15$ for Filament 2 indicates a density (n$_{\rm e}$ $\leq$ 200
cm$^{-3}$) and close to the low density limit of the [\ion{S}{2}] line ratio. 
Electron densities at or near the low density [\ion{S}{2}] ratio
limit of 1.4 is typical for SNRs \citep{Fesen85}. The S/N of the other spectra
where [\ion{S}{2}] line emission is seen is too low to give reliable
density estimates.

\subsection{Shock Emission Filaments}

Balmer dominated emission filaments like those seen for Filament 1
have been seen in a few Galactic SNRs and are thought to be the result of a
high velocity shock moving through a partially neutral medium producing Balmer
dominated emission filaments \citep{CKR80}. The H$\alpha$ emission line profile
for such filaments consists of two components; a narrow component generated by
the post-shock electron excitation of cold preshock neutral hydrogen atoms
passing through a collisionless shock front, and a much broader component
created via charge exchange between neutral hydrogen atoms and hot postshock
protons. 

Such Balmer filaments are characteristically thin and mark
the current location of a remnant's advancing interstellar shock front.  They
are sometimes referred to as nonradiative filaments to distinguish them from
the typically brighter, cooling radiative shock filaments commonly seen in
SNRs.  Examples of Balmer dominated filaments include the optical filaments in
the Tycho and SN 1006 remnants, portions of the Kepler and RCW 86 remnants, and
an array of faint, thin outlying filaments along the eastern and northern
boundaries of the Cygnus Loop (see \citealt{Heng10} for a review).

In young remnants like Tycho, Kepler, and SN 1006, the estimated shock velocity
responsible for Balmer dominated filaments is $\simeq 1500 - 2000$ km s$^{-1}$.
However, much lower shock velocities in the range of $150 - 450$ km s$^{-1}$
have been estimated for Balmer dominated filaments in the Cygnus Loop
\citep{Hester94,Medina14} and in G159.6+7.3 \citep{FM10}. 

The optical filaments seen in G70 are similar in morphology to Balmer filaments
in the the Cygnus Loop \citep{Hester94,Medina14} and G156.2+5.7 \citep{GF07}.
Specifically, G70 filaments seen in Regions 1, 2 (Fig.\ \ref{Reg1n5}) and 
parts of Region 3 appear strikingly similar to Balmer filaments seen
along the outer northeastern limb of the Cygnus Loop.  

While the spectrum for Filament 1 indicates a Balmer dominated
shock emission, the presence of other emission lines in the spectrum for
Filament 2 are unusual for Balmer dominated filaments.  Faint
[\ion{N}{2}] and [\ion{S}{2}] line emissions have been seen in some Balmer
filaments in the Cygnus Loop \citep{FI85}, Kepler's SNR \citep{Blair91}, and
Tycho's SNR \citep{Ghav00} and have been attributed to shock precursor
emissions ionizing the preshock material located out ahead of the shock front
\citep{Hester94,Ghav00}.  

Some other Balmer dominated filaments show weak [\ion{O}{3}] 5007 \AA \ line
emission.  For example, in the case of Cygnus Loop Balmer filaments,
[\ion{O}{3}] 5007 \AA \ has been observed with a strength around 10\% that of
H$\beta$ and can be even stronger in cases where the shocked filament is
undergoing a transition from a nonradiative adiabatic shock to one that is
radiative \citep{Hester94}.  Our detection of [\ion{O}{3}] emission for
Filament 4 is of similar strength relative H$\alpha$ to these other
cases. 

However, the presence of [\ion{N}{1}] 5198, 5200 \AA, [\ion{O}{1}] 6300, 6364
\AA \ line emissions like seen in the Region 3 and 5 spectra has never been
reported in any Balmer dominated filament spectrum but are common in radiative
SNR filaments. Although weak [\ion{N}{1}] and [\ion{O}{1}] emissions could be
the result of shock precursor emission, we suspect it is more likely to be due
to a Balmer dominated filament transitioning from nonradiative to radiative.
The presence of [\ion{O}{3}] emission positioned some distance behind the
leading edge of Region 5's filament fits with the picture of gas
transitioning from nonradiative to radiative emission.

\begin{figure*}[t]
\epsscale{0.8}
\centering
\plotone{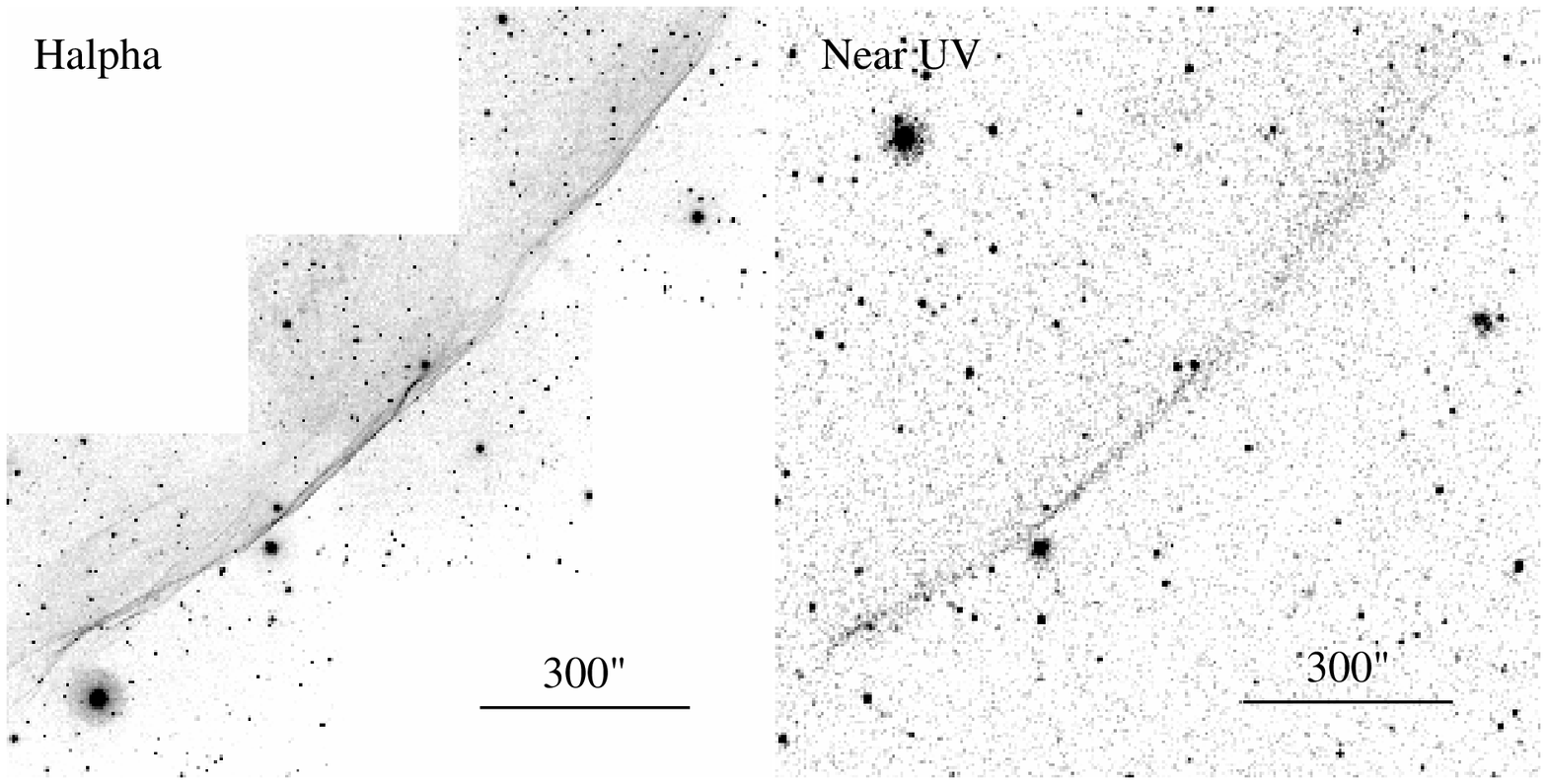}
\plotone{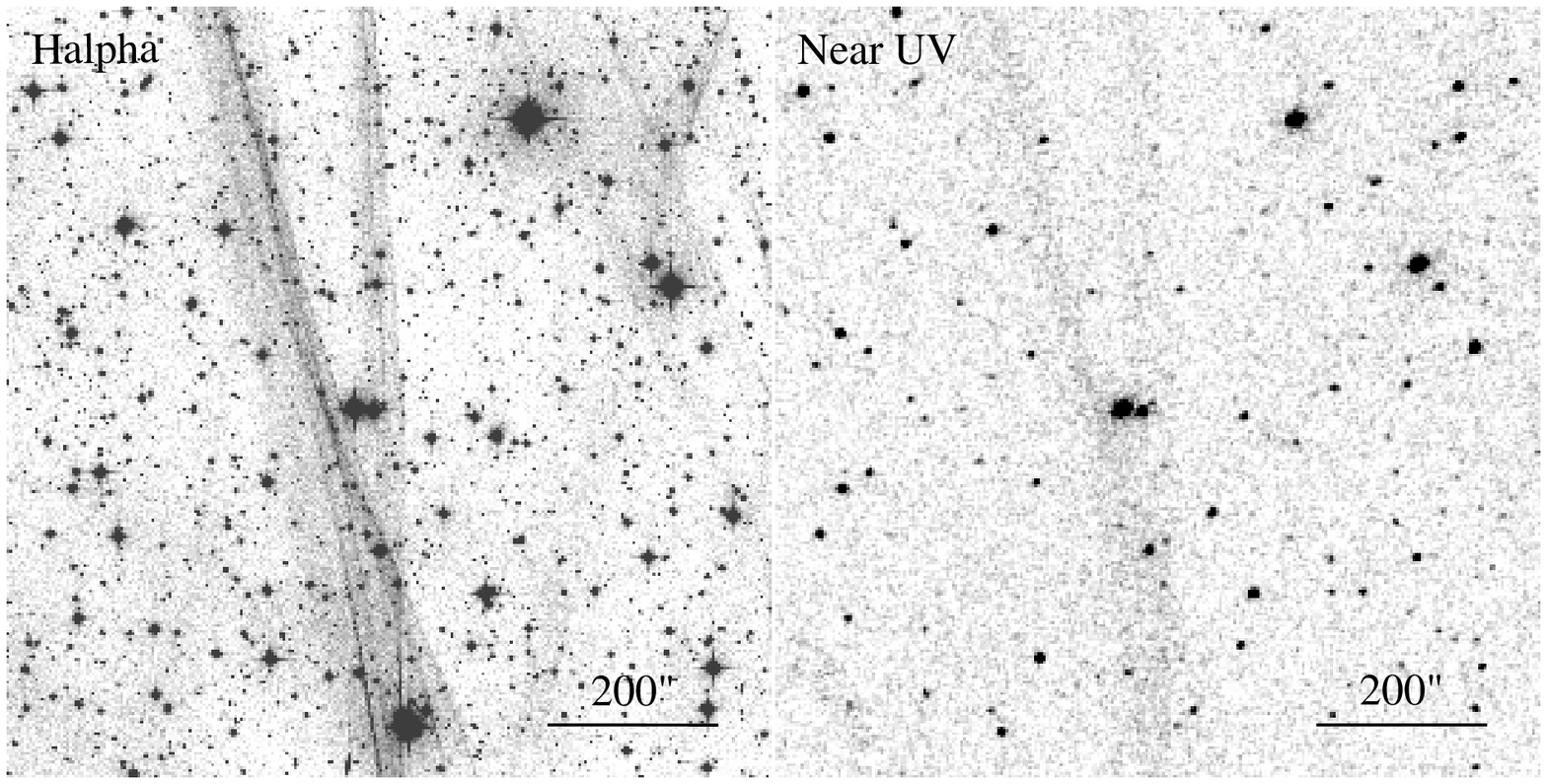}
\plotone{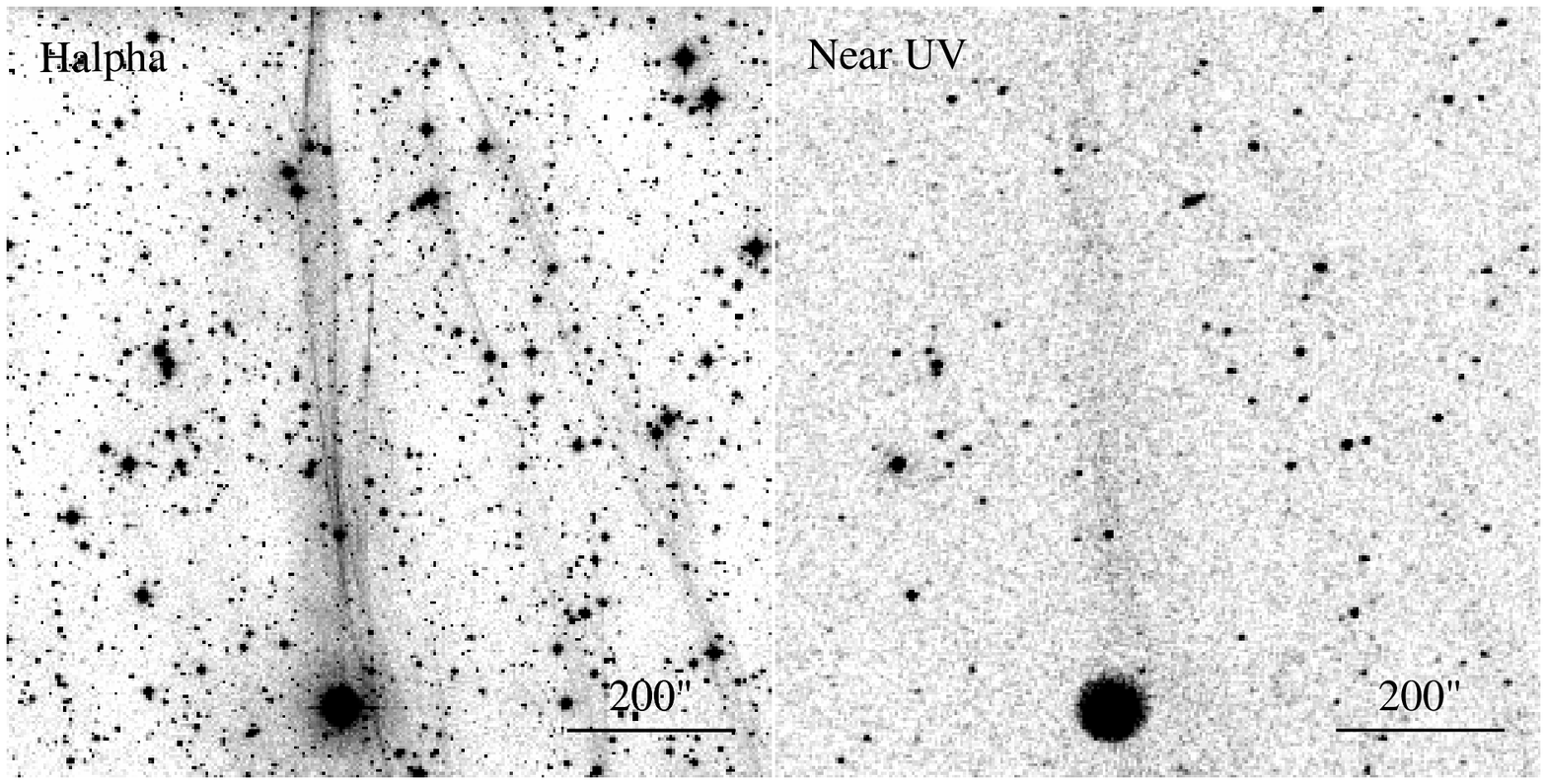}
\caption{Comparison of H$\alpha$ and {\sl GALEX} near UV emission
($1770 -  2730 $ \AA) images for Regions 3, 6, and 7 of
$G70.0-21.5$ in the top, middle, and bottom panels, respectively.
}
\label{UVimages}
\end{figure*}

Shocks below 90 km s$^{-1}$ are predicted to exhibit 
strong line emissions from neutral and singly ionized species such as
[\ion{N}{1}], [\ion{N}{2}], [\ion{S}{2}] and [\ion{O}{1}] with weak or absent
[\ion{O}{3}] 5007 emission \citep{Shull79,Raymond79,Dopita84}. Thus, a Balmer
dominated spectrum plus a contribution from a radiative $75 - 90$ km s$^{-1}$
shock like that in Model C of \citet{Shull79} could generate a spectrum
consistent with that observed for Regions 3 and 5.  Support for this
scenario is the presence of relatively strong [\ion{O}{1}] 6300, 6364 \AA \
line emission ([\ion{O}{1}] 6300/H$\alpha$ $> 0.1$) which is commonly seen in
SNRs and is characteristic of radiative shock emission \citep{Fesen85}.

Additional evidence for shock velocities below 100 km s$^{-1}$ for G70
filaments comes from the detection of faint near UV emission seen on {\sl
GALEX} images.  As shown in Figure \ref{UVimages}, several of G70's brighter
filaments exhibit emission within {\sl GALEX's} 1770 - 2730 \AA \ near UV
bandpass presumably due to filament \ion{C}{3}] 1909 \AA \  line emission. Far
UV {\sl GALEX} images ($1350 - 1780$ \AA) covering these same filaments show
only exceedingly weak emission presumably from \ion{C}{4} 1550 \AA.  Shock
models predict weaker \ion{C}{4} than \ion{C}{3}] in shocks $\leq$ 90 km
s$^{-1}$ \citep{Shull79,Raymond79}. Thus, a large \ion{C}{3}] 1909/\ion{C}{4}
1550 emission ratio is consistent with a relatively low shock velocity as
suggested by the presence of the [\ion{N}{1}] and [\ion{O}{1}] emissions seen
in Filament 2.


\begin{figure*}ht]
\centering
\plotone{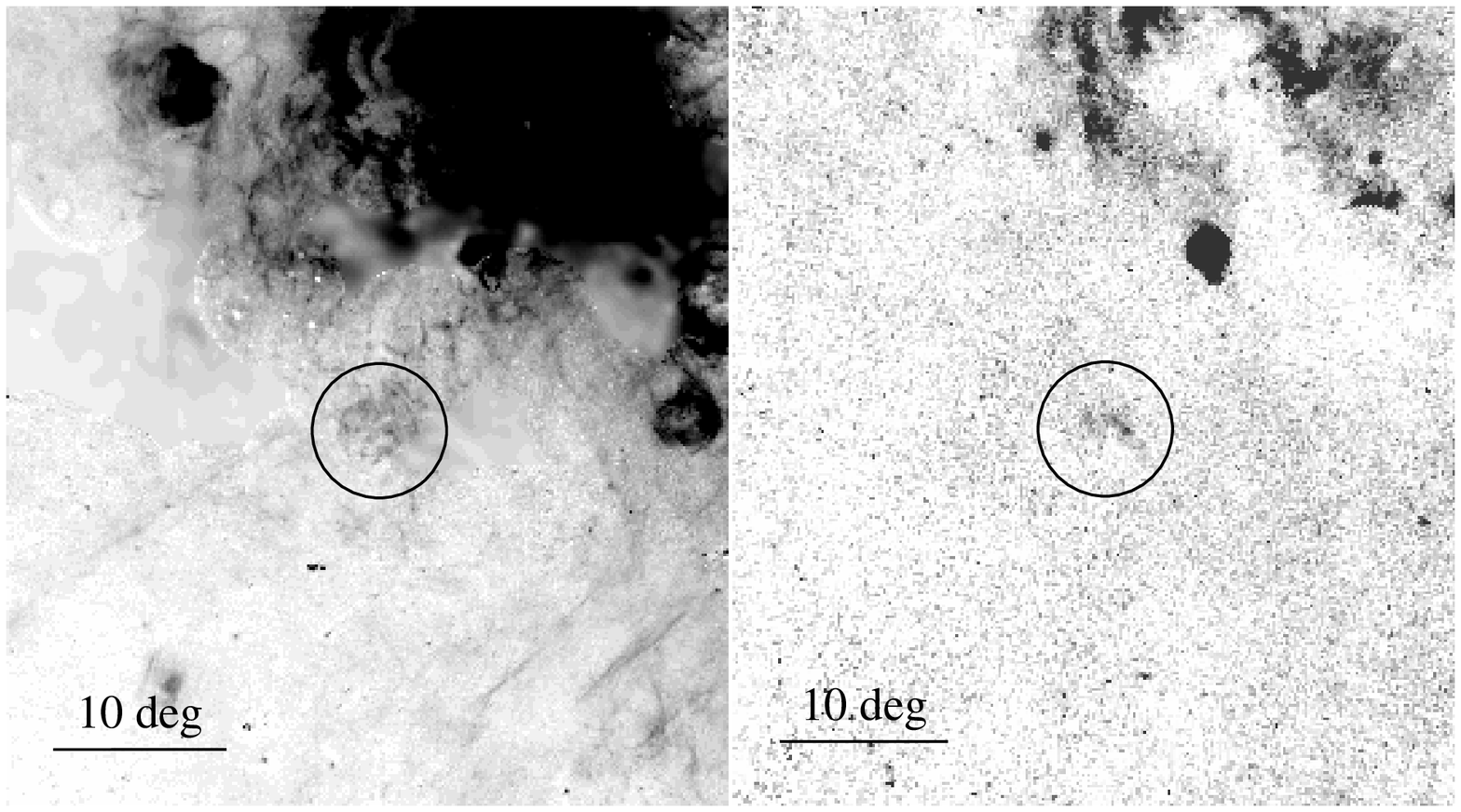}
\plotone{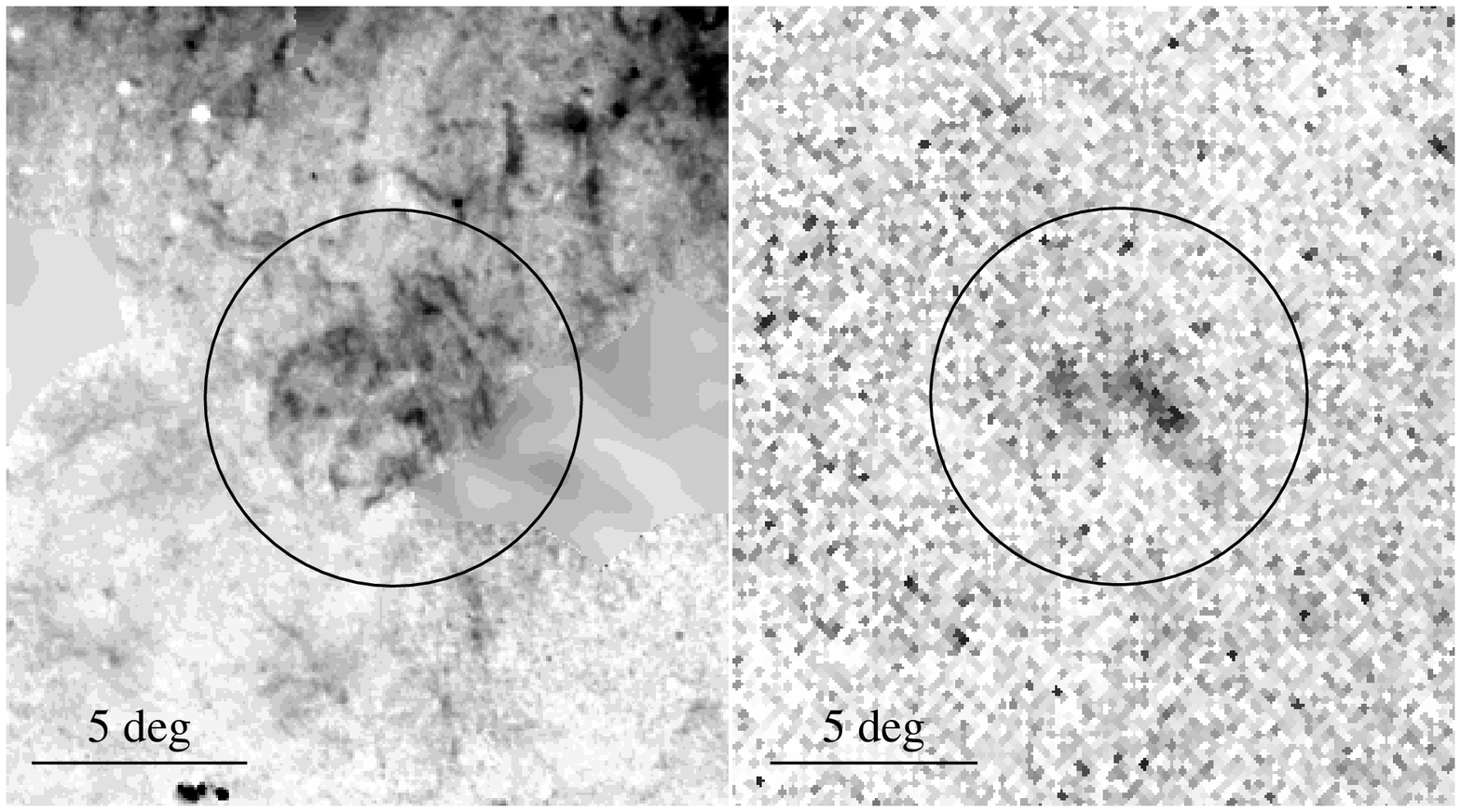}
\caption{Upper panels show the VTSS H$\alpha$ image (left) and the
{\sl ROSAT} All Sky Background Survey (RASS3) 
X-ray image (right) of a large region off the Galactic plane centered
on G70.0-21.5 (marked by the circles) illustrating the lack of X-ray emission
off the galactic plane around G70.0-21.5.
North is up, East to the left.
The remnant's associated X-ray emission
is curved and lies on the remnant's side closest to the Galactic plane.
The bright elliptical X-ray source 12 degrees northwest of G70.0-21.5 is the Cygnus Loop SNR.
Lower panels show enlargements of these H$\alpha$ and X-ray images. }
\label{Xray}
\end{figure*}

\begin{figure*}[ht]
\centering
\plotone{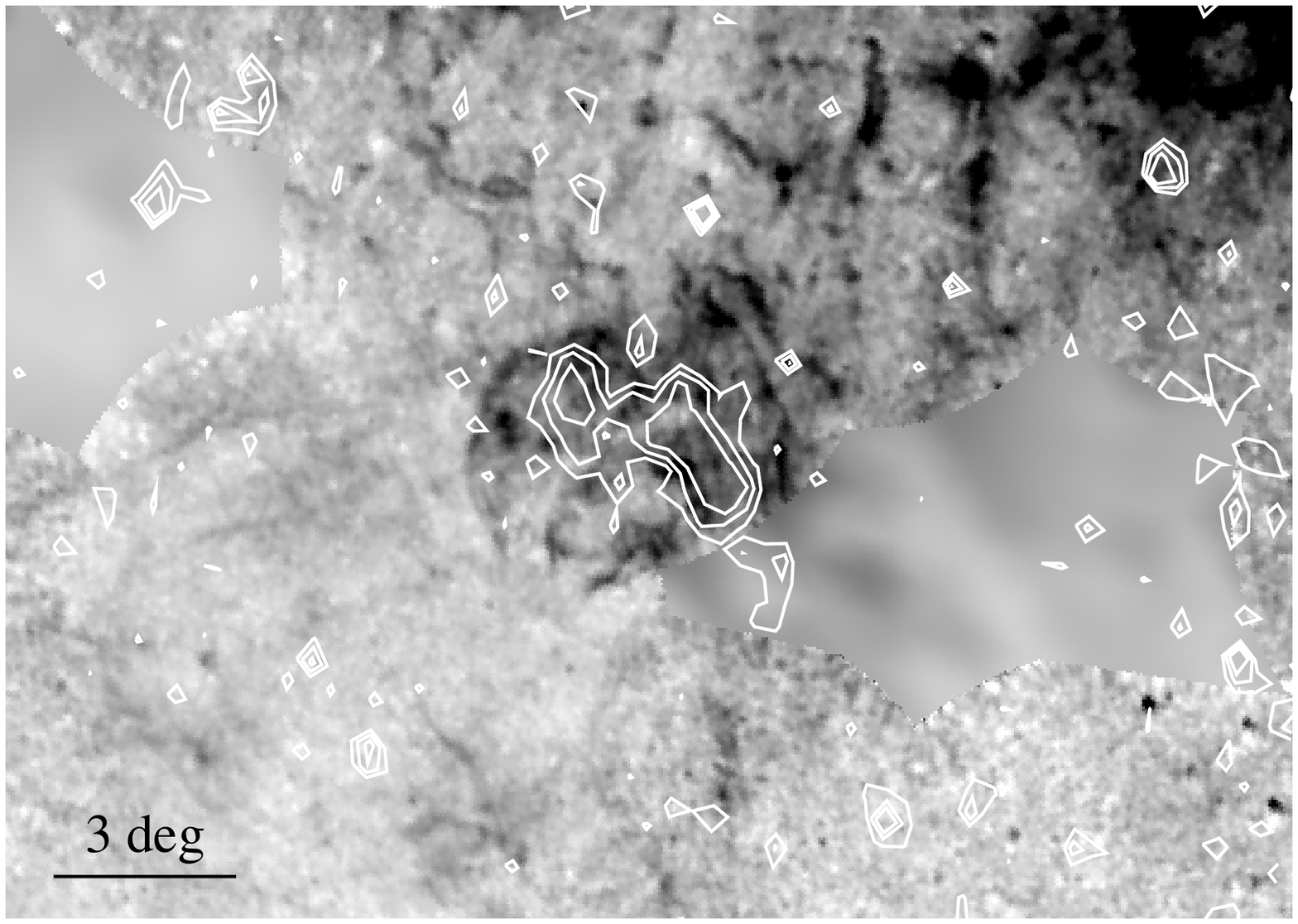}
\caption{X-ray emission contours from the {\sl ROSAT} All Sky Background Survey (RASS3) 
onto the VTSS H$\alpha$ image of G70.0-21.5.
North is up, East to the left.
Note that the detected X-ray emission lies within the projected 
boundaries of the remnant's optical emission.}
\label{Xray_overlay}
\end{figure*}

\subsection{G70.0-21.5: A New Galactic Supernova Remnant}

We propose that the large G70.0-21.5 emission shell is a previously
unrecognized Galactic supernova remnant.  Both the morphology and spectra of
G70.0-21.5's optical filaments are consistent with Balmer dominated shock
emission transitioning into radiative filaments.  The presence of diffuse,
extended X-ray emission coincident with the G70 nebula lends further support
to a supernova remnant identification.  

Figure \ref{Xray} shows VTSS H$\alpha$ and {\sl ROSAT} X-ray images of the G70
region.  A faint, diffuse, crescent shaped X-ray emission nebula can be seen to
be coincident with the portion of the H$\alpha$ nebulosity closest to the
Galactic plane as shown in the optical and X-ray overlay shown in Figure
\ref{Xray_overlay}.  The presence of this isolated patch of X-ray emission
situated far off the galactic plane and coincident with G70's location is
consistent with a SNR nature.

The location of this X-ray emission projected inside the remnant's optical
emission boundaries is also consistent with that of an evolved SNR where the
X-ray emission appears centrally filled \citep{Shelton99,Williams04}.  Although
the X-ray flux from this emission patch is too weak to produce a meaningful
X-ray spectrum, its positional coincidence, size, and alignment with G70's
northern shell lends additional support to G70's SNR identification.

The lack of reported nonthermal radio emission at G70's location
\citep{Shimmins68,Green14} is not surprising given the nebula's location of
more than 21 degrees off the Galactic plane and thus outside the galactic
latitude range usually searched for nonthermal radio emission (e.g., the
Canadian Galactic Plane Survey; \citealt{Taylor03}).  Of the 294 Galactic SNRs
cataloged, G70.0-21.5 lies the furthest off the plane of the Galaxy, some 6
degrees farther than the current record-holder, the Lupus Loop (G330.0+15.0),
and over 12 degrees farther off the plane than any SNR in the first quadrant
\citep{Green14b}.

Although it does lie within the regions covered in galactic H~I surveys aimed
at identifying large shell-like H~I structures, no H I shell possibly
associated with G70 has been reported \citep{Ehlerov13,Sallmen15}.  This
may be due in part to such H I surveys being particularly sensitive to
large, low velocity old ISM shells with typical ages greater than several
10$^{6}$ yrs \citep{Ehlerov13}, and thus much older than G70 and other large but
optical and X-ray emitting SNRs. 

With a diameter of roughly $4.0\degr \times 5.5\degr$,
G70 ranks among the largest galactic SNRs known in terms of angular size.
Because of its size and the presence of diffuse emission patches along its
western limb, it is not known whether it is still in the Sedov-Taylor
adiabatic expansion phase or has entered the later radiative expansion
phase. Consequently, many of its basic physical parameters (age, distance,
shock velocity, local ISM density) are uncertain.  We can, however, place rough
upper limits to its maximum distance based on arguments regarding its
likely maximum physical size and z-distance off the galactic plane.

Based on a shock velocity estimate from our optical spectra, distances around 1
-- 2 kpc are likely from comparisons with other large remnants.  At a
distance of only 1 kpc, G70's linear dimensions would be  $70 \times 95$ pc,
already larger than almost all known SNRs. Conversely, at distances less than 1 kpc, G70
would be among the closest remnants. Compared to the Cygnus Loop which has an
estimated shock velocity of $150 - 350$ km s$^{-1}$ and diameter of $30 - 60$
pc (\citealt{Blair09,Medina14}) and the Monoceros Loop's v = 50 km s$^{-1}$
(dia.\ $\simeq$100 pc; \citealt{Leahy86,Odegard86}), a velocity $\simeq$75 km
s$^{-1}$ for G70 suggests it is less evolved than Monoceros but older than the
Cygnus Loop and hence of a size intermediate to these.

G70 is unlikely to lie at distances much greater than 3 kpc in order to
have a z-distance less than the 500 pc scale-height of the local H I halo
including the Local Spur \citep{Lockman86,Cer09} and the z-distance of most
metal poor stars \citep{Bovy12}.  Conversely, distances around 1.0 kpc or less
raises the possibility of detecting motion of G70's sharp filaments.  

As mentioned in $\S$3.1, some of G70's filaments are visible on the red Palomar
Sky Survey images and we examined and compared our 2014 images with the
digitized 1954 and 1991 POSS images (DSS1 and DSS2) covering the G70 region.
G70 filaments were too weakly detected on the DSS1 images to be useful, but
comparisons of our 2014 and the DSS2 1991 images for G70's brighter filaments
indicated proper motions in the range $0\farcs03 - 0\farcs06$ yr$^{-1}$.
Although these values have significant measurement uncertainties due to the very weak
detection of the filaments on the broadband DSS2 images, such proper motions
are consistent with a distance around 1 kpc assuming a shock velocity
$\simeq$75 km s$^{-1}$.         
 
Lastly, follow-up studies of this apparent SNR should prove fruitful.
Detection of nonthermal radio emission from G70 could firmly establish its SNR
nature.  However, its radio emission might be quite weak, as SNR evolutionary
models predict a rapid drop in radio emission from large remnants
\citep{Asvarov06}.  Furthermore, observations of extragalactic SNRs suggest
especially weak radio emission for SNRs expanding in low density media
\citep{Pannuti02}.  Nonetheless, existing or future galactic radio surveys
(e.g., the Arecibo G-ALFA Transit Survey; \citealt{Taylor10}) might detect
nonthermal emission from G70.

The probable low ambient density at G70's high z-distance is likely the cause
for its optical emission to be nearly entirely Balmer dominated filaments, a
property rare in SNRs.  Although faint, the presence of so many Balmer
dominated filaments throughout G70 and the variation of emission line strength
we observed across some of its filaments may make it a valuable remnant for
better understanding the transition of nonradiative to radiative shocks.
Finally, investigation of its X-ray emission could provide 
information regarding its shock velocity and age.  

In summary, we believe the nebulosity, G70.0-21.5, to be a physically large and
relatively old SNR that lies unusually high off the galactic plane. It exhibits
a rich array of beautiful and overlapping Balmer dominated filaments
covering a region over 4 degrees in size. At a likely distance around 1 to 2 kpc,
G70.0-21.5 also ranks among the closest remnants known. 

\bigskip

We thank an anonymous referee for comments and a careful reading that improved
the paper's content and presentation, and D. Patnaude and D. Milisavljevic for
helpful discussions.

\begin{deluxetable}{cccccc}
\tablecolumns{6}
\tablecaption{Imaged Regions of G70.0-21.5}
\tablewidth{0pt}
\tablehead{
\colhead{Region} &
\colhead{RA (J2000)} &
\colhead{Dec (J2000)} &
\colhead{ $l$ } &
\colhead{ $b$ } &
\colhead{Exposures} \\
\colhead{ } &
\colhead{ h \ m \ s} &
\colhead{~~~  \degr ~~ \arcmin ~~ \arcsec} &
\colhead{~  \degr ~~ \arcmin ~~ \arcsec} &
\colhead{~  \degr ~~ \arcmin ~~ \arcsec} &
\colhead{(s)}
}
\startdata
1 & 21:35:18 & +19:19:54 & 71:55:03 & -23:33:17 & $3 \times 600$ \\
2 & 21:35:56 & +18:36:04 & 71:26:32 & -24:09:40 & $3 \times 600$ \\
3 & 21:25:08 & +16:57:07 & 68:13:46 & -23:20:08 & $2 \times 600$ \\
4 & 21:29:21 & +17:15:26 & 69:12:08 & -23:53:49 & $2 \times 600$ \\
5 & 21:21:59 & +17:26:56 & 68:06:40 & -22:26:30 & $2 \times 600$ \\
6 & 21:13:51 & +19:04:38 & 68:07:04 & -19:55:01 & $ 3 \times 600$\\
7 & 21:14:10 & +19:26:37 & 68:28:05 & -19:44:20 & $3 \times 900$ \\
8 & 21:15:16 & +20:28:34 & 69:29:00 & -19:16:23 & $1 \times 900$ \\
9 & 21:19:50 & +21:29:09 & 71:01:50 & -19:46:31 & $2 \times 900$ \\
\enddata
\end{deluxetable}


\begin{thebibliography}{}
\bibitem[Asvarov(2006)]{Asvarov06} Asvarov, A.~I.\ 2006, \aap, 459, 519
\bibitem[Blair et al.(2009)]{Blair09} Blair, W.~P., Sankrit, 
         R., Torres, S.~I., Chayer, P., \& Danforth, C.~W.\ 2009, \apj, 692, 335 
\bibitem[Blair et al.(1991)]{Blair91} Blair, W.~P., Long, K.~S., \& Vancura, O.\ 1991, \apj, 366, 484 
\bibitem[Bovy et al.(2012)]{Bovy12} Bovy, J., Rix, H.-W., \& Hogg, D.~W.\ 2012, \apj, 751, 131 
\bibitem[Chevalier et al.(1980)]{CKR80} Chevalier, R.~A., 
         Kirshner, R.~P., \& Raymond, J.~C.\ 1980, \apj, 235, 186 
\bibitem[Cersosimo et al.(2009)]{Cer09} Cersosimo, J.~C., Muller, R.~J., \& Figueroa, N.~S.\ 2009, \apj, 699, 716 
\bibitem[de Gasperin et al.(2014)]{deGasp14} de Gasperin, F., Evoli, C., Br{\"u}ggen, M., et al.\ 2014, \aap, 568, A107
\bibitem[Dennison et al.(1998)]{Dennison98} Dennison, B., Simonetti, J.~H., \& Topasna, G.~A.\ 1998, 
         Publications of the Astronomical Society of Australia, 15, 147
\bibitem[Dopita et al.(1984)]{Dopita84} Dopita, M.~A., Binette, L., Dodorico, S., 
         \& Benvenuti, P.\ 1984, \apj, 276, 653 
\bibitem[Downes(1971)]{Downes71} Downes, D.\ 1971, \aj, 76, 305 
\bibitem[Drew et al.(2005)]{Drew05} Drew, J.~E., Greimel, R., Irwin, M.~J., et al.\ 2005, \mnras, 362, 753 
\bibitem[Ehlerov{\'a} \& Palou{\v s}(2013)]{Ehlerov13} Ehlerov{\'a}, S., \& Palou{\v s}, J.\ 2013, \aap, 550, A23 
\bibitem[Fesen et al.(1982)]{Fesen82} Fesen, R.~A., Blair, W.~P., \& Kirshner, R.~P.\ 1982, \apj, 262, 171 
\bibitem[Fesen et al.(1985)]{Fesen85} Fesen, R.~A., Blair, W.~P., \& Kirshner, R.~P.\ 1985, \apj, 292, 29 
\bibitem[Fesen \& Itoh(1985)]{FI85} Fesen, R.~A., \& Itoh, H.\ 1985, \apj, 295, 43 
\bibitem[Fesen \& Milisavljevic(2010)]{FM10} Fesen, R.~A., \& Milisavljevic, D.\ 2010, \aj, 140, 1163 
\bibitem[Finkbeiner(2003)]{Fink03} Finkbeiner, D.~P.\ 2003, ApJS, 146, 407
\bibitem[Gao \& Han(2014)]{Gao14} Gao, X.~Y., \& Han, J.~L.\ 2014, \aap, 567, A59 
\bibitem[Gerardy \& Fesen(2007)]{GF07} Gerardy, C.~L., \& Fesen, R.~A.\ 2007, \mnras, 376, 929 
\bibitem[Gerbrandt et al.(2014)]{Gerbrandt14} Gerbrandt, S., Foster, T.~J., 
         Kothes, R., Geisb{\"u}sch, J., \& Tung, A.\ 2014, \aap, 566, 76 
\bibitem[Ghavamian et al.(2000)]{Ghav00} Ghavamian, P., Raymond, J., Hartigan, P., \& Blair, W.~P.\ 2000, \apj, 535, 266 
\bibitem[Gonz{\'a}lez-Solares et al.(2008)]{GS08} Gonz{\'a}lez-Solares, E.~A., 
         Walton, N.~A., Greimel, R., et al.\ 2008, \mnras, 388, 89 
\bibitem[Green(2014a)]{Green14} Green, D.~A.\ 2014, Bulletin of the Astronomical Society of India, 42, 47 
\bibitem[Green(2014b)]{Green14b} Green, D.~A.\ 2014, IAU Symposium, 296, 188
\bibitem[Gull et al.(1977)]{Gull77} Gull, T.~R., Kirshner, R.~P., \& Parker, 
         R.~A.~R.\ 1977, \apjl, 215, L69 
\bibitem[Heng(2010)]{Heng10} Heng, K.\ 2010, Pub.\ Astro.\ Soc.\ Pacific, 27, 23 
\bibitem[Hester et al.(1994)]{Hester94} Hester, J.~J., Raymond, J.~C., \& Blair, W.~P.\ 1994, \apj, 420, 721 
\bibitem[Lockman et al.(1986)]{Lockman86} Lockman, F.~J., Hobbs, L.~M., \& Shull, J.~M.\ 1986, \apj, 301, 380 
\bibitem[Leahy et al.(1986)]{Leahy86} Leahy, D.~A., Naranan, S., \& Singh, K.~P.\ 1986, \mnras, 220, 501 
\bibitem[Mar{\v s}{\'a}lkov{\'a}(1974)]{Mar74} Mar{\v s}{\'a}lkov{\'a}, P.\ 1974, \apss, 27, 3 
\bibitem[Massey \& Gronwald(1990)]{Massey90} Massey, P., \& Gronwald, C. 1990, \apj, 358, 344
\bibitem[Medina et al.(2014)]{Medina14} Medina, A.~A., Raymond, J.~C., Edgar, R.~J., et al.\ 2014, \apj, 791, 30 
\bibitem[Milne(1970)]{Milne70} Milne, D.~K.\ 1970, Australian Journal of Physics, 23, 425 
\bibitem[Neckel \& Vehrenberg(1985)]{Neckel85} Neckel, T., \& Vehrenberg, H.\ 1985, Atlas of Galactic Nebulae, 
         Vol.\ I \& II, Duesseldorf: Treugesell-Verlag, 1985
\bibitem[Odegard(1986)]{Odegard86} Odegard, N.\ 1986, \apj, 301, 813 
\bibitem[Oke(1974)]{Oke74} Oke, J. B.  1974, \apjs, 27, 21
\bibitem[Osterbrock \& Ferland(2006)]{Oster06} Osterbrock, D.~E., \& Ferland, G.~J.\ 2006, 
         Astrophysics of gaseous nebulae and active galactic nuclei, 2nd.~ed.~Sausalito, CA: University Science Books  
\bibitem[Pannuti et al.(2002)]{Pannuti02} Pannuti, T.~G., Duric, N., Lacey, C.~K.,
         Ferguson, A.~M.~N., Magnor, M.~A., \& Mendelowitz, C.\ 2002, \apj, 565, 966
\bibitem[Parker(1967)]{Parker67} Parker, R.~A.~R.\ 1967, \apj, 149, 363 
\bibitem[Parker et al.(1979)]{Parker79} Parker, R.~A.~R., Gull, T.~R., \& Kirshner, R.~P.\ 1979, NASA SP-434  
\bibitem[Raymond(1979)]{Raymond79} Raymond, J.~C.\ 1979, \apjs, 39, 1 
\bibitem[Raymond et al.(1980)]{Raymond80} Raymond, J.~C., Davis, 
         M., Gull, T.~R., \& Parker, R.~A.~R.\ 1980, \apjl, 238, L21 
\bibitem[Renaud et al.(2010)]{Renaud10} Renaud, M., Marandon, V., 
         Gotthelf, E.~V., et al.\ 2010, \apj, 716, 663 
\bibitem[Reynolds et al.(2013)]{Reynolds13} Reynolds M. T., et al., 2013, \apj, 766, 112
\bibitem[Sabin et al.(2013)]{Sabin13} Sabin, L., Parker, Q.~A., Contreras, M.~E., et al.\ 2013, \mnras, 431, 279 
\bibitem[Sallmen et al.(2015)]{Sallmen15} Sallmen, S.~M., Korpela, E.~J., Bellehumeur, B., et al.\ 2015, \aj, 149, 189 
\bibitem[Sankrit et al.(2005)]{Sankrit05} Sankrit, R., Blair, 
         W.~P., Delaney, T., et al.\ 2005, Advances in Space Research, 35, 1027
\bibitem[Sharpless(1959)]{Sharpless59} Sharpless, S.\ 1959, \apjs, 4, 257 
\bibitem[Shimmins \& Day(1968)]{Shimmins68} Shimmins, A.~J., \& Day, G.~A.\ 1968, Australian Journal of Physics, 21, 377 
\bibitem[Shelton(1999)]{Shelton99} Shelton, R.~L.\ 1999, \apj, 521, 217
\bibitem[Shull \& McKee(1979)]{Shull79} Shull, J.~M., \& McKee, C.~F.\ 1979, \apj, 227, 131 
\bibitem[Sivan(1974)]{Sivan74} Sivan, J.~P.\ 1974, \aaps, 16, 163
\bibitem[Stupar et al.(2008)]{SP08} Stupar, M., Parker, Q.~A., \& Filipovi{\'c}, M.~D.\ 2008, \mnras, 390, 1037 
\bibitem[Taylor et al.(2003)]{Taylor03} Taylor, A.~R., Gibson, S.~J., 
         Peracaula, M., et al.\ 2003, \aj, 125, 3145 
\bibitem[Taylor \& Salter(2010)]{Taylor10} Taylor, A.~R., \& Salter, C.~J.\ 2010, 
        The Dynamic Interstellar Medium: A Celebration of the Canadian Galactic Plane Survey, 438, 402 
\bibitem[Williams et al.(2004)]{Williams04} Williams, R.~M., Chu, Y.-H., Dickel, J.~R.,
          Gruendl, R.~A., Shelton, R., Points, S.~D., \& Smith, R.~C.\ 2004, \apj, 613, 948
\end{thebibliography}
\end{document}